\documentclass{elsart3p}
\usepackage{graphicx}
\usepackage{amssymb}
\begin{document}

\begin{frontmatter}

\title{
Vortex Transport and Pinning Effectiveness in Conformal Pinning Arrays
} 
\author{
D. Ray$^{1,2}$, C. Reichhardt$^2$, C. J. Olson Reichhardt$^2$, 
and B. Jank{\' o}$^{1}$  
} 
\address{
$^1$Department of Physics, University of Notre Dame, Notre Dame,
Indiana 46556, USA\\ 
$^2$Theoretical Division,
Los Alamos National Laboratory, Los Alamos, New Mexico 87545, USA
} 

\begin{abstract}
We examine the current driven dynamics for vortices interacting with 
conformal crystal pinning arrays
and compare to the dynamics of vortices driven over random pinning 
arrays. We find that the pinning is enhanced in the conformal
arrays 
over a wide range of fields, consistent with previous results 
from flux gradient-driven simulations. 
At 
fields above this range,  
the effectiveness of the
pinning in the moving vortex state 
can be  
enhanced 
in the random arrays compared to the conformal arrays, leading to 
crossing of the velocity-force curves.       
\end{abstract}
\end{frontmatter}

\section{Introduction}
Many of the applications of type-II superconductors require that the 
system maintain a large critical current 
or effective pinning of vortices 
in the presence of a magnetic field \cite{1}. 
One approach to this 
problem  
is the use of 
lithography 
to create 
arrays of artificial pinning sites
\cite{2,3,4,5,6} 
by means of 
nanohole lattices 
\cite{2,3,4,5} or 
arrays of magnetic dots \cite{6}. 
This raises the question of what arrangement of pinning sites 
maximizes the effectiveness of the pinning, for a given number of sites. 
In periodic arrays of pinning sites, 
strong commensuration or matching 
effects can occur when the number of vortices equals an integer multiple $n$
of the number of pinning sites \cite{2,3,7,8}. At the matching conditions, 
there can be a peak in the critical current when the vortices 
form an ordered state \cite{3,7,8}.  
The enhancement of pinning at commensurate fields has also been 
observed in colloidal experiments \cite{9} on periodic optical
trap arrays.
The colloids are repulsively interacting particles that have behavior
similar to that of vortices, showing that 
understanding
vortex dynamics on periodic or semi-periodic substrates is also useful for 
the general understanding of dynamics near commensurate-incommensurate 
transitions \cite{10}.   
For fields close to matching fields,
interstitials or vacancies can appear in the ordered vortex structure and
act as effective particles that are 
weakly pinned \cite{11}. 
However, away from the matching fields, the critical current falls off 
substantially and the pinning becomes less effective. 

Other approaches to pinning enhancement include the use of quasiperiodic
pinning arrays such as Penrose tilings \cite{12,13,14}. 
Commensuration effects still occur for such arrays;
moreover, 
the strength of the pinning at nonmatching fields is generally 
improved from 
that found for nonmatching fields in periodic or random
pinning arrays
\cite{12,13}. 
The random dilution of periodic pinning arrays produces peaks
in the critical current not only at the matching fields 
but also at fields
where the number of vortices matches the number of pinning sites
in the original undiluted array \cite{15}. 
Strong non-integer matching peaks in the critical current also appear
in honeycomb pinning arrays, which are an example of an ordered diluted 
triangular pinning array \cite{16}. 
Enhancement of the pinning at fractional fields can be 
achieved using artificial ice pinning array geometries \cite{17},
and these fractional matching peaks can be as strong as or even stronger 
than the integer matching peaks \cite{18}. 

\begin{figure}
\begin{center}
\includegraphics[width=\columnwidth]{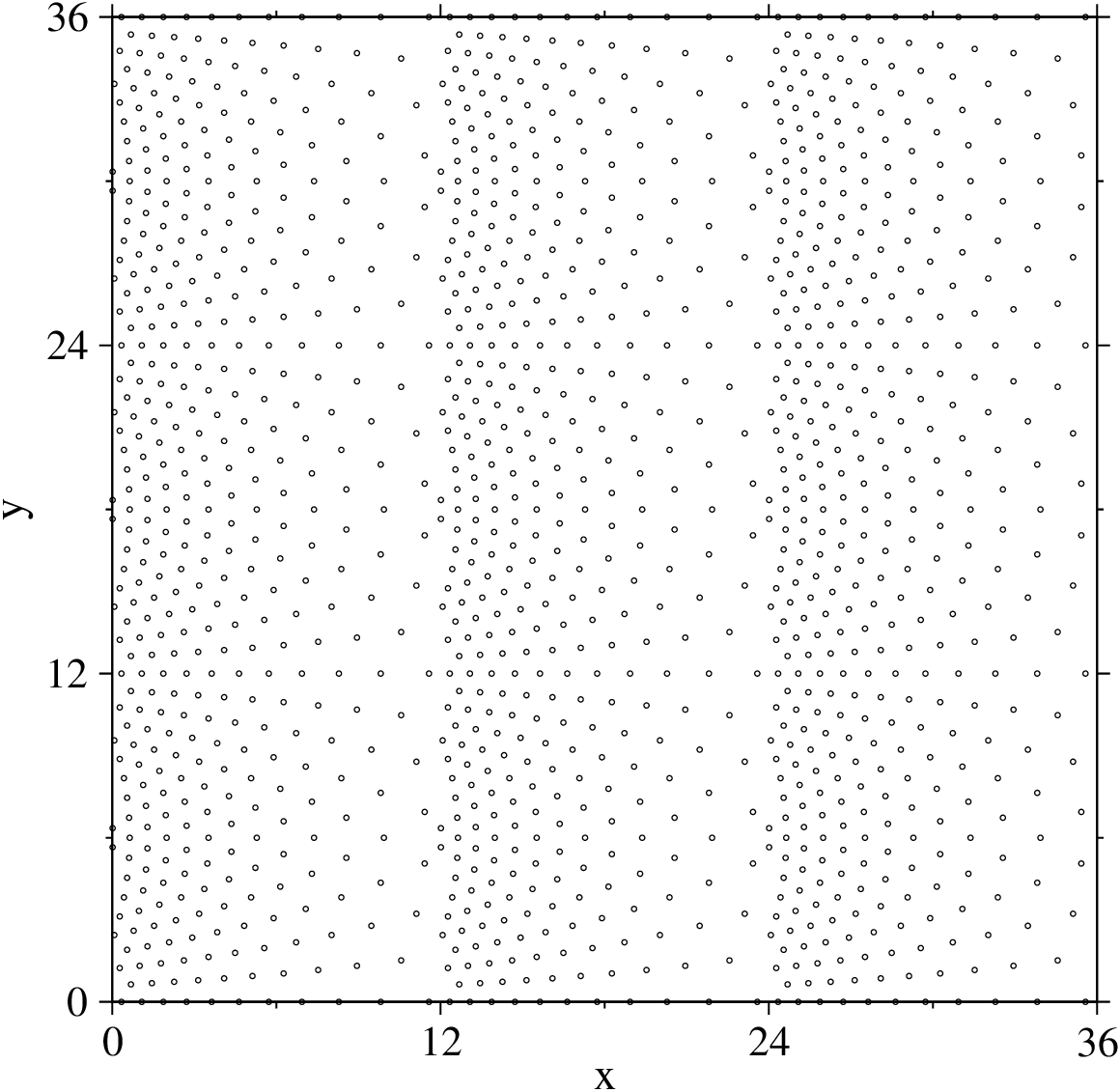}
\end{center}
\caption{
The conformal pinning geometry used for 
simulated transport measurements.
A random pinning array with the same number of pins is used for 
comparison (not shown).
Periodic boundary conditions apply in both the $x$- and $y$-directions. 
The drive is applied in the positive $x$-direction.
 }
\label{fig:1}
\end{figure}

All the previously mentioned approaches have the deficiency of 
exhibiting reduced pinning effectiveness away from 
certain special field values. 
To address this issue, 
a new type of pinning geometry called a conformal crystal pinning
array (illustrated in Fig.~1) was recently proposed \cite{19}.
This array 
is constructed by performing a 
conformal transformation \cite{20} on a triangular lattice 
to create a gradient in the pinning density while preserving the
local sixfold ordering of the original triangular array. 
Flux gradient simulations show that the overall critical current 
in the conformal
pinning array (CPA) is
enhanced over that of a uniform random pinning 
array \cite{19} 
for a wide range of fields, 
and is also higher than that of uniform periodic
pinning arrays except for fields very close to integer matching, where
the periodic pinning gives a marginally higher critical current. 
The gradient in pinning density present in the CPA enhances
the pinning since it can match the gradient in the vortex
density, 
and also leads to an absence of commensuration effects 
or peaks in the critical current. 
Random pinning arrays with a density gradient equivalent to that
of the CPA
produce a small critical current enhancement 
compared to uniform random arrays; however, a CPA with the same number
of pinning sites gives a substantially larger critical current,
indicating that the preservation of the
sixfold ordering of the pinning array is also important for 
enhancing the pinning \cite{19}. 
The simulation predictions were subsequently confirmed in 
experiments which compared CPAs to
random and periodic arrays \cite{21,22}. Other work on non-conformal pinning arrays containing gradients includes
numerical studies of hyperbolic tesselations \cite{23}, as well as
experiments on non-conformal pinning arrays with gradients
in which the pinning was enhanced compared to uniform arrays \cite{24}. 

The first numerical work on CPAs focused on flux-gradient driven simulations 
where the critical current is proportional to the width of the 
magnetization loop \cite{19}.  In such simulations, there is a gradient
in the vortex density across the sample \cite{FLUXGRAD}.
One question is whether the CPA
still produces enhanced pinning
in systems driven with an applied current. 
Previous work indicated that the CPA produces
a pinning enhancement compared to random arrays in this case as well \cite{19}.
In this work we further explore the current-driven system
by varying the applied magnetic field
and analyzing the vortex velocity as a function of external drive 
to produce a measurement that is
proportional to an experimentally measurable voltage-current curve. 
We find that at very low vortex densities, the difference in critical
current between random arrays
and CPAs is small, and that as the field increases, 
the conformal arrays have stronger effective pinning, producing both
a larger depinning force and a lower average vortex velocity in the 
moving state compared to random arrays. 
At higher fields, the CPA still has a high depinning threshold; 
however, once the vortices are in the moving 
state, the average vortex velocity for the random arrays can be 
lower than that for the CPA, indicating that the effectiveness of the pinning
in the dynamic regime is suppressed for the CPA compared to the
random pinning.
We show that this arises due to the earlier onset of 
dynamical ordering \cite{25,26} in CPAs compared to random pinning
arrays at these higher magnetic fields.

\section{Simulations}   
We consider  an effective 2D model of pointlike vortices
where a single vortex $i$ obeys the following equation of motion:
\begin{equation}  
\eta \frac{d {\bf R}_{i}}{dt} = 
{\bf F}^{vv}_{i} +  {\bf F}^{P}_{i} +  {\bf F}^{D}_{i}. 
\end{equation} 
Here $\eta=\phi_0^2d/2\pi\xi^2\rho_N$ is the damping constant, 
$d$ is the sample thickness, $\phi_0 = h/2e$ is the elementary flux quantum, 
and $\rho_{N}$ is the normal-state resistivity of the material; 
we work in units where $\eta$ is set equal to 1.
Vortex $i$ is located at ${\bf R}_i$. 
The vortex-vortex repulsive interaction force is 
${\bf F}^{vv}_{i}= \sum^{N_{v}}_{j\neq i}F_{0}K_1(R_{ij}/\lambda){\hat {\bf R}}_{ij}$, 
where $K_{1}$ is the modified Bessel function, 
$\lambda$ is the London penetration depth, 
$F_{0} = \phi_0^{2}/(2\pi\mu_{0}\lambda^{3})$,
$R_{ij} = |{\bf R}_i - {\bf R}_{j}|$ 
is the distance between vortex $i$ and vortex $j$, and the unit vector
${\hat {\bf R}}_{ij} = ({\bf R}_{i} - {\bf R}_{j})/R_{ij}$.    
The force from the pinning sites is given by ${\bf F}^{P}$.
Various models for the pinning can be considered; here, we use
parabolic attractive sites with
\begin{equation}
{\bf F}^{P}_{i} = -\sum^{N_{p}}_{k=1}(F_{p}/r_{p})({\bf R}_{i} - {\bf R}^{(p)}_{k})
\Theta[r_{p} - |{\bf R}_{i} - {\bf R}^{(p)}_{k}|],  
\end{equation}
where ${\bf R}^{(p)}_{k}$ is the location of pinning site $k$, 
$F_{p}$ is the maximum pinning force, 
$r_{p}$ is the pinning radius,  
and $\Theta$ is the Heaviside step function.  
Finally, an externally applied current ${\bf J}$ produces a Lorentz force 
${\bf F}^D_i={\bf J} \times {\bf B}$ on the vortices 
that is perpendicular to the applied current. 

To measure the transport properties of a given pinning array, 
we first place $N_v$ vortices randomly and allow them to anneal; 
then we 
apply 
a slowly increasing  
driving force ${\bf F}^D=F_d{\bf \hat x}$
in the $x$ direction and measure the 
resulting average 
vortex velocity in the $x$ direction,  
$\langle v_{x}\rangle = (1/N_{v})\sum^{N_{v}}_{i=1}
{\bf v}_{i}\cdot {\hat {\bf x}}$, 
where ${\bf v}_{i} = d{\bf R}_{i}/dt$.
The system geometry is illustrated in Fig.~1, 
where we show a conformal pinning array. The construction of the 
CPA is described in \cite{19}. 
Transport measurements were also performed with an array of 
randomly distributed pinning sites for comparison. 
In both arrays, the pinning density is maintained at $n_p=1.0/{\lambda}^2$ 
with pinning radius $r_p=0.12\lambda$ and pinning force 
$F_p=0.55F_0$. 
Simulations performed with these parameter choices should be 
in the same regime as 
recently conducted experiments on CPAs \cite{21,22}.
The system size is $36\lambda\times 36\lambda$, 
with periodic boundary conditions in the
$x$- and $y$-directions.  
In this work, we characterize transport as a function of the number 
of vortices $N_v$; we report this number as $B/B_{\phi}=N_{v}/1296$, 
where $B$ is the average magnetic field in the sample resulting from 
the vortices and $B_{\phi}$ is the matching field achieved when the number of 
vortices equals the number of pinning sites.

\section{Transport} 

\begin{figure}
\begin{center}
\includegraphics[width=\columnwidth]{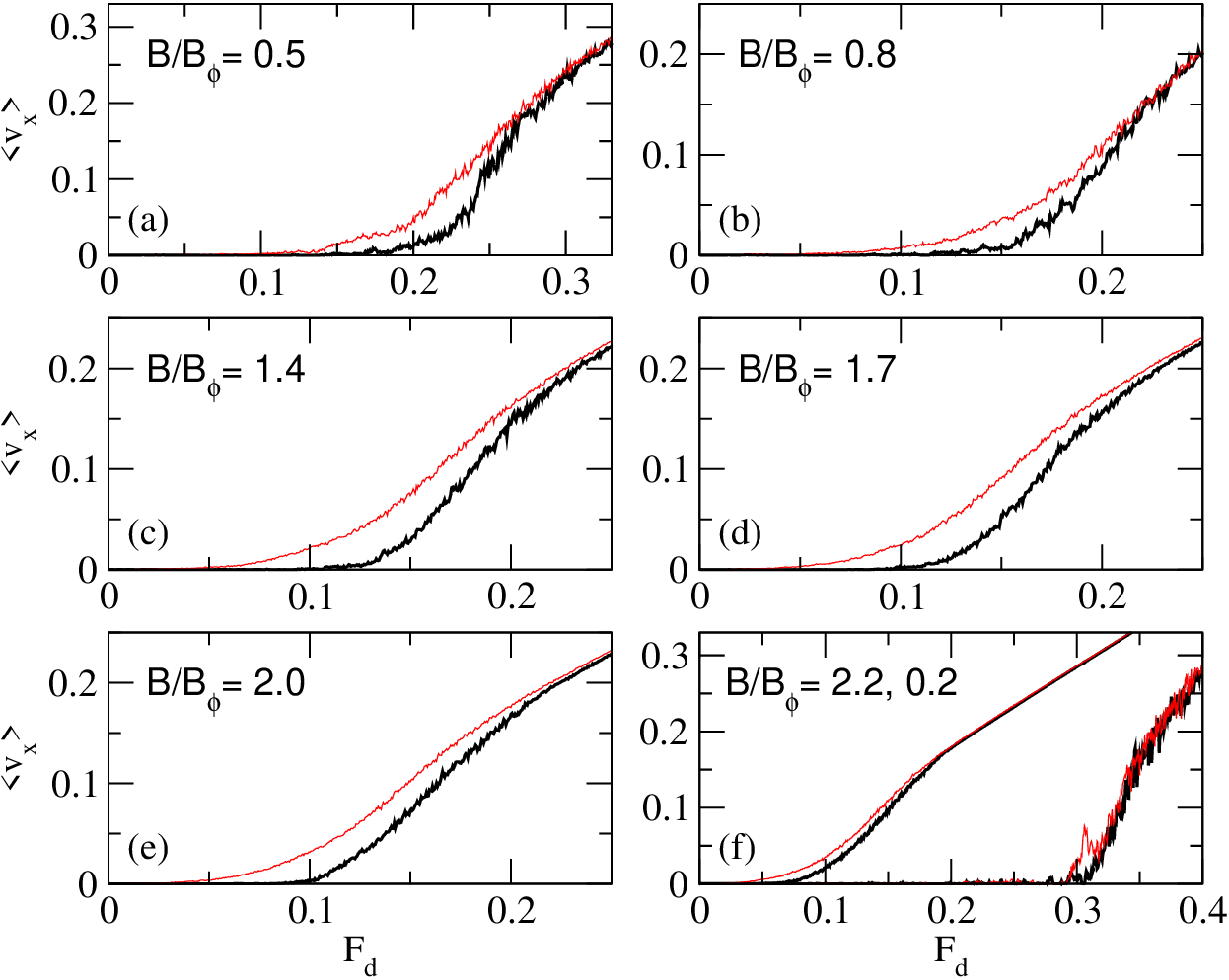}
\end{center}
\caption 
{$\langle v_x\rangle$ versus $F_d$ curves for the CPA (lower dark lines)
and random arrays (upper light lines) 
for $B/B_\phi=$ (a) 0.5, (b) 0.8, (c) 1.4, (d) 1.7, (e) 2.0, and
(f) 0.2 (right lines) and 2.2 (left lines).
Panels (a-e) show that 
the effectiveness of the pinning for the CPA is higher than for the
random array over a wide range of fields. 
Panel (f) shows field levels at the extremes of this range, where the 
CPA is no longer more effective than random pinning.    
} 
\label{fig:2}
\end{figure}

\begin{figure}
\includegraphics[width=\columnwidth]{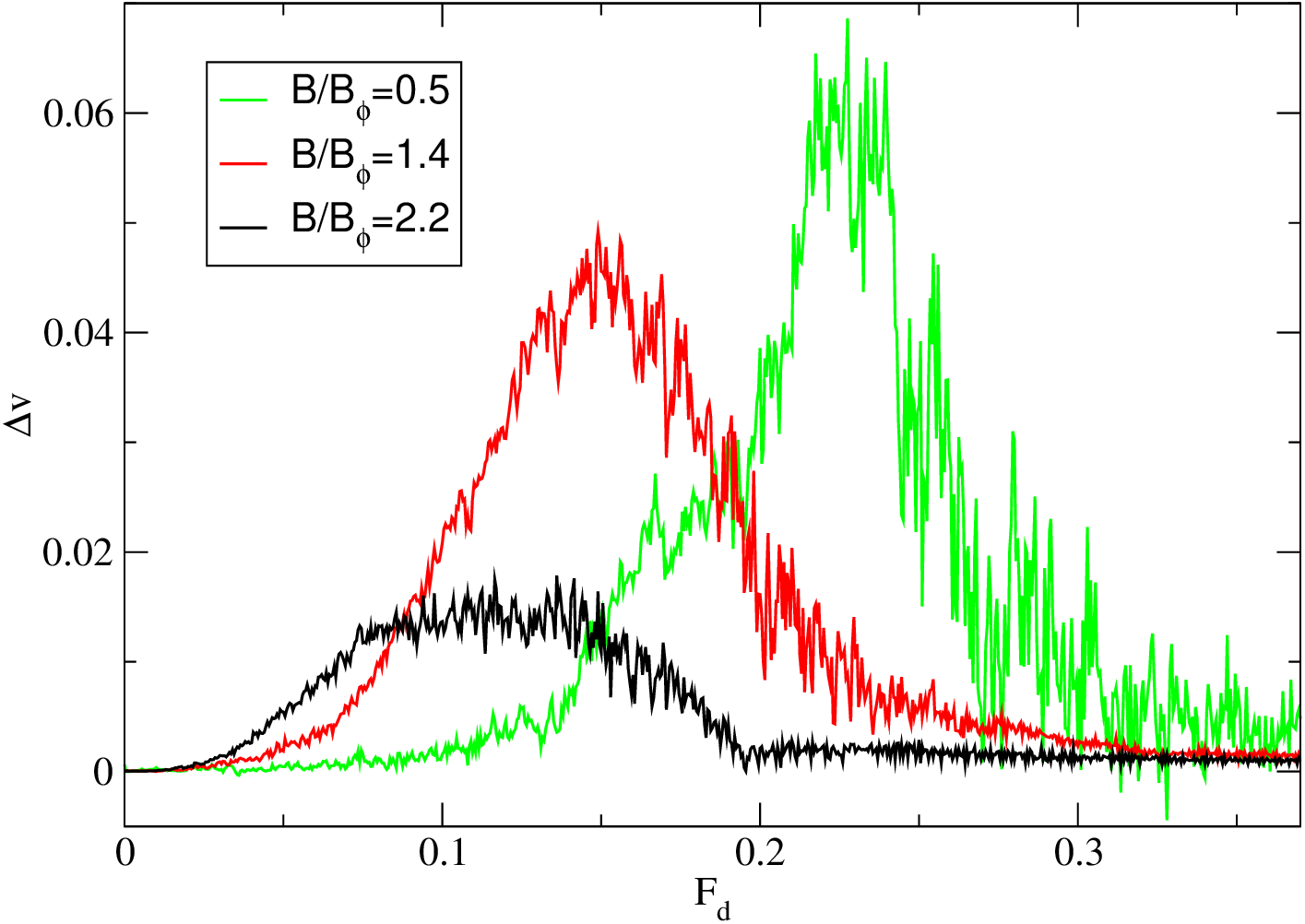}
\caption
{
The difference between the velocity response in the random array and
the CPA, $\Delta v = \langle v_x^{\rm rand}\rangle - \langle v_x^{\rm CPA}\rangle$,
vs $F_d$ for samples with $B/B_\phi=$ 0.5, 1.4, and 2.2 (upper right to 
lower left).
In this field range, $\Delta v$ is positive, indicating that
the pinning is more effective in the CPA at low and
intermediate values of $F_d$.
At the highest drives, 
the response of both arrays becomes Ohmic and 
$\Delta v$ goes to zero.
}     
\label{fig:3}
\end{figure}

In Fig.~2 
we plot $\langle v_{x}\rangle$ versus $F_{d}$ 
for uniform random arrays and CPAs at fields ranging
from $B/B_\phi=0.2$ to 2.2. 
Panels (a-e) show that 
over a wide range of fields from 0.5 to 2.0, vortices consistently 
remain stationary or pinned up to a higher drive in the CPA than 
in the random array, 
providing evidence that  
the pinning in the CPA 
is more effective than in the random array. 
This range of fields where we see increased CPA effectiveness in transport 
simulations is consistent with the corresponding range found for  
magnetization using quasistatic flux-gradient driven simulations in 
\cite{19}.
Moreover, even above the depinning threshold, 
vortices continue to move more slowly through the CPA;   
we show this explicitly in Fig.~3 where 
we plot the difference between the velocity response in the random array
and the CPA,
$\Delta v =  \langle v^{\rm rand}_{x}\rangle - 
\langle v^{\rm CPA}_{x}\rangle$, as a function of $F_d$.
We see that the pinning in the CPA is more effective
than in the random array, with a positive $\Delta v$ for all but the 
highest values of $F_d$.

In panel (f) of Fig.~2, we explore field values at the edges of the 
range, where the CPA loses its increased effectiveness. At a low field 
level, $B/B_{\phi} = 0.2$, the transport curves lie on top of each other. 
Because there are so few vortices present in the system, only a small 
percentage of the pinning sites in an array are actually pinning vortices 
at any given time. The pinning arrays are being very sparsely sampled, 
and so the details of their structure do not come into play. 
Conversely, at a high field level $B/B_{\phi} = 2.2$, the dense packing 
of vortices in the system begins to overwhelm the pinning. 
The CPA has a spatially varying density of pinning sites; in the 
gradient-driven simulations of \cite{19}, it was shown that the CPA 
begins to fail when the vortex density exceeds the maximum density of 
pinning sites, which occurs at one edge of the CPA (corresponding to 
$x=0\lambda, 12\lambda, 24\lambda$ in Fig.~1). The CPA used in this work 
has a maximum pinning density of $2.0{\lambda}^{-2}$, so this explanation 
is consistent with our results.

Fig.~3 brings out this trend of decreased CPA effectiveness at high fields. 
For example, we can examine the maximum $\Delta v$ achieved: 
at $B/B_{\phi} = 1.4$, $\Delta v$ reaches a maximum value of 0.045,
while for $B/B_{\phi}=2.2$ the maximum value of $\Delta v$ is only 0.015,
indicating that as $B/B_{\phi}$ increases, the difference in 
pinning effectiveness of the CPA compared to the random pinning array 
is reduced.  
We can also look at the transition to the Ohmic regime which occurs 
for large $F_{d}$, 
where $\Delta v$ goes to 0
as 
all the vortices flow freely in response to the large driving force, 
and the effects of the pinning become minimal. 
The transition to the Ohmic response regime
occurs near $F_{d} = 0.26$ for $B/B_{\phi} = 1.4$ in Fig.~3, while
for $B/B_{\phi} = 2.2$, the transition 
drops to a lower value of 
$F_{d} = 0.19$.

\section{CPA Breakdown}

\begin{figure}
\includegraphics[width=\columnwidth]{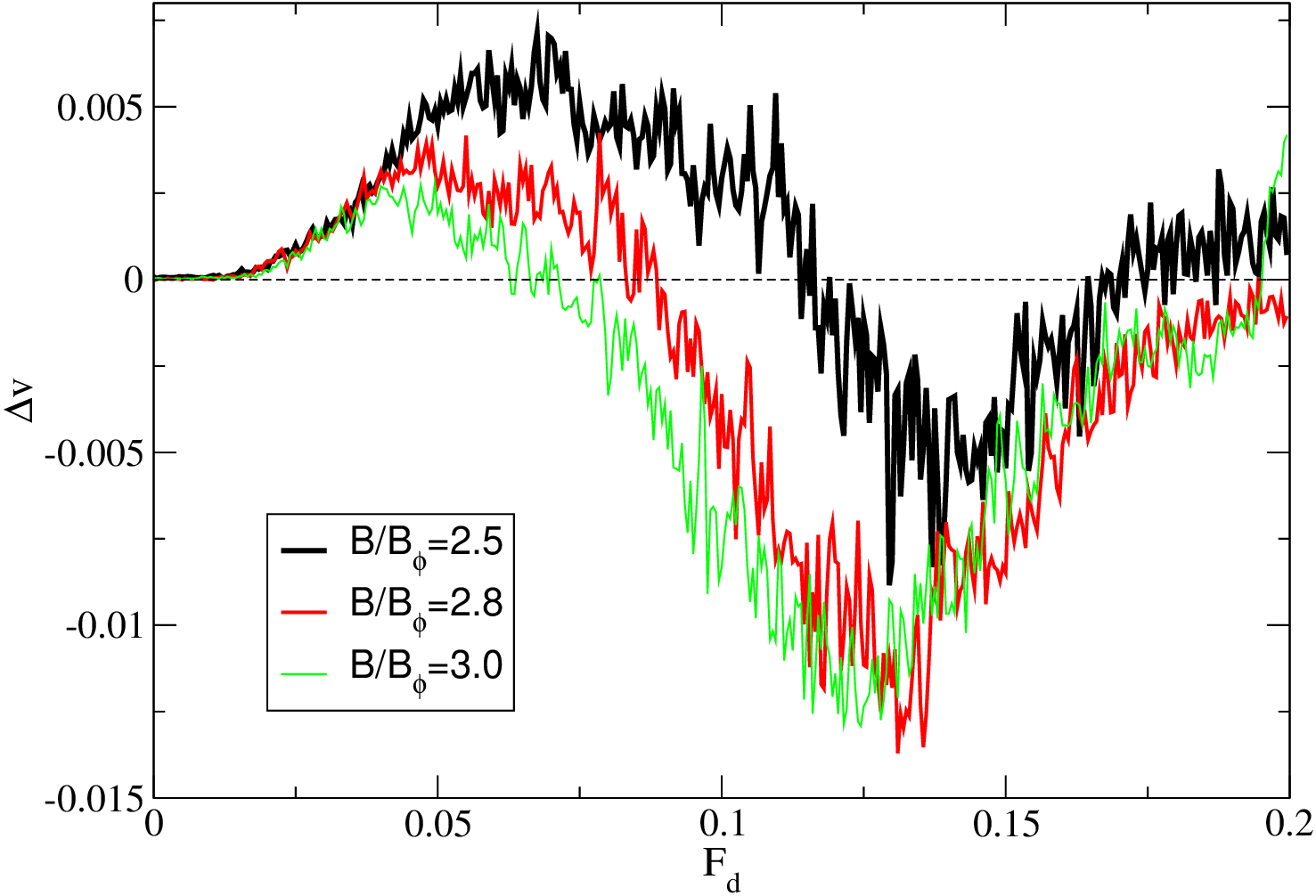}
\caption
{
$\Delta v$ vs $F_d$ for samples with high field values of 
$B/B_\phi=2.5$, 2.8, and 3.0 (upper right to lower left).
For low $F_d$, the pinning is more effective in the CPA, as indicated
by the positive value of $\Delta v$; however, 
at intermediate $F_d$,  
the pinning becomes more effective in the random 
arrays 
as shown by the negative $\Delta v$.
}
\label{fig:4}
\end{figure}

If we consider even higher field values, then a new feature arises in 
the velocity-force curves. In Fig.~4 we plot the quantity $\Delta v$ 
defined in the previous section for $B/B_{\phi} = 2.5$, 2.8, and $3.0$. 
For each of these fields $\Delta v$ is initially positive, but drops
below zero as $F_d$ increases, 
indicating that the average vortex velocity is higher
in the CPA than in the random array at intermediate values of $F_{d}$, 
so that CPA pinning actually becomes less effective than random pinning.
The reversal of the effectiveness of the pinning in the moving state produces
a crossing in the velocity vs force curves, as shown
in panel (c) of Fig.~5 
($B/B_{\phi} = 2.8$) 
compared to panels (a-b)
($B/B_{\phi} = 1.7$, 2.2).

The reversal of the effectiveness in the pinning at 
intermediate drives occurs because the vortices dynamically order or
partially crystallize
at a lower drive
in the CPA than in the
random pinning array. 
It is known from current-driven simulation studies of random pinning arrays 
that  
a dynamical reordering transition can occur into a moving state that is 
partially crystalline or smectic-like \cite{25,26}.
In the dynamically ordered state, the vortex velocities are generally
higher than in moving states with more random ordering,
since the shear modulus of a random structure is much lower.
A disordered vortex configuration
has a higher probability of some vortices
being temporarily pinned by the substrate,
while in a moving crystal state, the vortices all move together and
can not be individually trapped by pinning sites.
For the random array, as the field increases, the drive $F_d^{Or}$ 
at which the vortices begin to dynamically order decreases. 
$F_d^{Or}$ is also a function of the pinning density $n_p$, 
and as $n_{p}$ decreases, $F^{Or}_{d}$ also decreases. 
In the CPA, the pinning density has a gradient, and as a result, 
there is a gradient in the value of $F^{Or}_{d}$ across the sample. 
At the higher magnetic fields, the vortices can start to 
locally dynamically order in the lower pin density 
portions of the CPA sample. 

The partially ordered state forms in the low pin density regions during a
transient time $\tau_o$, and this state becomes disordered while passing
through the high pin density regions during a transient
time $\tau_d$.
As the field increases, these transient times change, 
and the vortices remain disordered 
if $\tau_{d} > \tau_{o}$, 
while for $\tau_{o} < \tau_{d}$  the vortices can order. 
This means that in a random pinning array, the vortices are
disordered when
$F_{d} < F^{Or}$; 
however, for a CPA at the same value of  $F_{d}$, 
if $\tau_{o} < \tau_{d}$, an ordered moving vortex state will form
and hence $v_{x}$ for the CPA will be higher than for the random array. 
As $B/B_{\phi}$ increases, $\tau_{o}$ decreases. 
This is consistent with the behavior in Fig.~4, where
the extent of the range of $F_d$ over which 
$\Delta v < 0$ 
grows as 
$B/B_{\phi}$ increases. 
It may be possible that at high enough $B/B_{\phi}$, 
the vortices in the CPA would immediately dynamically order as soon as
they depin;
in this case, the critical current for the random array 
would be higher than that of the CPA.

\begin{figure}
\begin{center}
\includegraphics[width=\columnwidth]{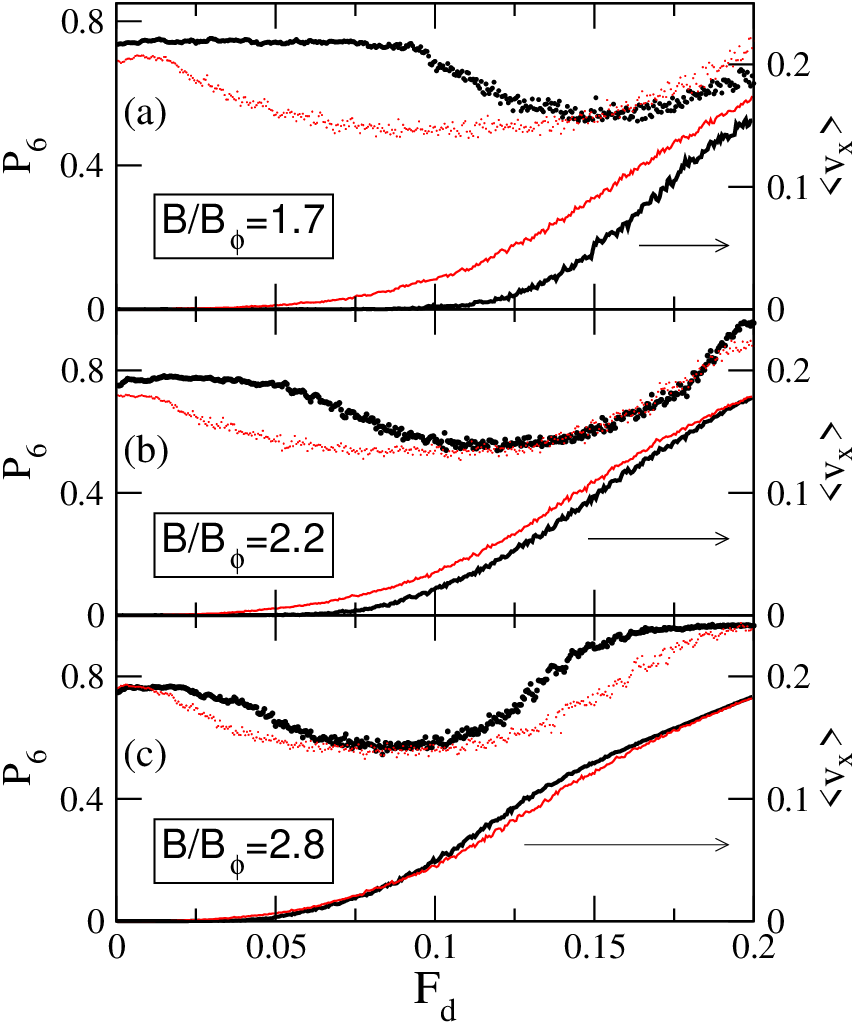}
\end{center}
\caption
{
Lower solid curves: $\langle V_{x}\rangle$ vs $F_d$
for a random array (light lines) 
and a CPA (dark lines).  Upper symbols: 
$P_6$, the fraction of sixfold coordinated vortices,
vs $F_{d}$ for a random array (light symbols) and a CPA (dark symbols).
(a) $B/B_{\phi} = 1.7$. (b) $B/B_{\phi} = 2.2$. 
(c) $B/B_{\phi} = 2.8$. 
In (c), the vortices
dynamically order at a lower drive for the CPA than 
for the random array, giving a lower value of $\langle V_{x}\rangle$ 
for the random array at the intermediate drives $0.1 < F_{d} < 0.2$.
}            
\label{fig:5}
\end{figure}

In Fig.~5 we plot simultaneously $\langle v_{x}\rangle$ 
and the fraction of six-fold coordinated vortices $P_{6}$ versus $F_{d}$ 
for the random pinning and the CPA.
In the dynamically ordered moving crystal state, $P_{6}$ is close to $1$
\cite{25,26}. 
In Fig.~5(a) at $B/B_{\phi} = 1.7$, the pinning is more effective in the 
CPA over the entire window of $F_d$ shown in the figure.
At depinning, $P_{6}$ drops for both types of pinning
as the system enters a plastic flow regime. 
At higher $F_d$, $P_{6}$ increases when the vortices begin
to reorder, and in Fig.~5(a), $P_6$ for the random array is higher than that
for the CPA  
for $F_{d} > 0.15$. 
In Fig.~5(b) for $B/B_{\phi} = 2.2$, we find a similar trend; 
however, in Fig.~5(c) for $B/B_{\phi} = 2.8$, 
$P_{6}$ is higher for the CPA than
for the random array for $F_{d} > 0.1$. 
This also corresponds to the range of $F_d$
over which $\langle v_{x}\rangle$ in the CPA is higher than in the 
random array. 
At $F_{d} = 0.2$, $P_{6}$ reaches nearly the same value for both arrays,
and the difference in $\langle v_x\rangle$ between the two arrays
also vanishes. This
result confirms that at high magnetic fields,
the vortices dynamically order at a lower drive for the CPA than for
a random pinning array.

\begin{figure}
\begin{center}
\includegraphics[width=\columnwidth]{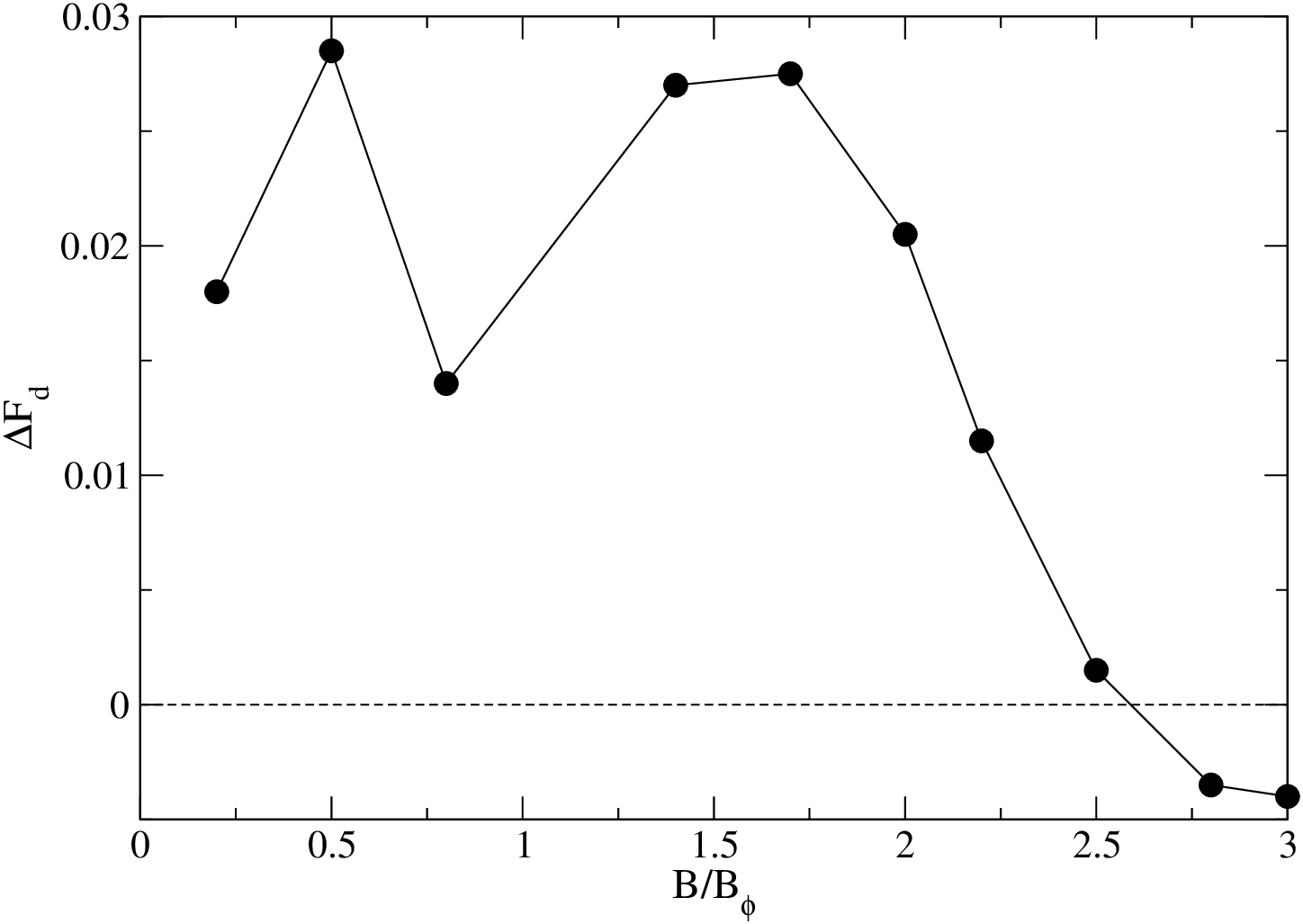}
\end{center}
\caption
{
The difference in the external drive $F_{d}$ at which 
$\langle v_x\rangle =0.05$ 
between the random and the conformal arrays, 
$\Delta F_d = F^{\rm rand}_{d}(\langle v_{x} \rangle = 0.05) 
- F^{\rm CPA}_{d}(\langle v_{x} \rangle =0.05)$, 
vs $B/B_{\phi}$. At low fields,
$\Delta F_d$
is small, at intermediate fields the CPA has stronger
pinning ($\Delta F_d>0$), and for $B/B_{\phi} > 2.5$ 
the random array has stronger pinning ($\Delta F_d<0$).  
}            
\label{fig:6}
\end{figure}

We can roughly estimate
the relative effectiveness of the pinning in the CPA and random pinning arrays
by plotting the difference in the value $F_{d}$ at which
$\langle v_{x} \rangle = 0.05$ for the two arrays,
$\Delta F_{d} =  F^{\rm rand}_{d}(\langle v_{x} \rangle = 0.05) 
- F^{\rm CPA}(\langle v_{x} \rangle = 0.05)$. 
Figure 6  
shows that
at low $B/B_{\phi}$, $\Delta F_d$ is small and
the difference between the random and conformal array is  
minimal due to the weak vortex-vortex interactions. 
$\Delta F_d$ is large and positive over the range 
$0.5 < B/B_{\phi} < 2.0$; 
it then decreases and 
becomes negative for $B/B_{\phi} > 2.5$.
While the exact value of $B/B_\phi$ at which $\Delta F_{d}$ drops below zero 
will depend on the velocity value chosen for the measurement, 
Fig.~6 
is consistent with the idea that the CPA is highly effective at fields 
less than the maximum local pinning density of the CPA,
but falls off in effectiveness above this value. 
It should be noted that 
the CPA effectiveness may 
also depend on the size $r_p$ and strength $F_p$ of the pinning sites, 
both of which can be sample dependent.

\section{Summary}

We investigated the current driven dynamics of vortices interacting 
with conformal pinning arrays and compared the effectiveness of the
pinning to that of random pinning arrays with the same total number 
of pinning sites. The conformal pinning array
is constructed by performing a conformal transformation of a 
triangular pinning array to create 
a new pinning array that has a density gradient but still conserves the 
local sixfold ordering of the original triangular array. 
We find that for vortex densities 
not exceeding 
the maximum local density of pinning sites in the conformal array, 
the critical depinning force for the
conformal array is higher than that of the random array; 
and furthermore, in the moving vortex state, the velocity of the vortices
in the conformal array is lower than in the random array.
At higher fields, the critical depinning force for the conformal array 
remains higher than that of the random array, but at intermediate drives
the average vortex velocity in the random arrays becomes lower 
than that in the conformal array, leading to a crossing of the velocity-force 
curves. 
This reversal of the pinning effectiveness 
arises because the vortices dynamically order at a lower drive in the 
conformal array than in the random array. 
Finally, at high drives, 
the difference between the two types of arrays is washed out
due to dynamical reordering of the vortices.

There are still issues to consider in the conformal pinning array, 
such as performing conformal transformations
on lattice structures other than a triangular array. 
It would also be interesting to investigate vortex ratchet effects of the
type previously found in samples with random or periodic pinning arrays
\cite{27,28}, as
these effects

This work was carried out under the auspices of the 
NNSA of the 
U.S. DoE
at 
LANL
under Contract No.
DE-AC52-06NA25396.

\end{document}